\documentclass[aps,pra,nofootinbib,showkeys,showpacs,amsfonts,amssymb,amsmath]{revtex4}

\usepackage{bm}
\usepackage[dvips]{epsfig}
\usepackage{color}
\definecolor{light-gray}{gray}{0.4}
\usepackage{amsmath}
\usepackage{amssymb}

\newcommand{\ket}[1]{|#1\rangle}
\newtheorem{dex}{Definition}
\newtheorem{example}{Example}[section]
\newtheorem{theorem}[example]{Theorem}
\begin{document}
\title{Degrees of entanglement for multipartite systems}%
\author{A. I. Solomon}
\email{a.i.solomon@open.ac.uk}
\affiliation{Department of Physics Sciences, The Open University,
Milton Keynes MK7 6AA, United Kingdom\\
and
\\
Laboratoire de Physique Th\'eorique de la Mati\`{e}re Condens\'{e}e (LPTMC),\\
Universit\'e Pierre et Marie Curie, CNRS UMR 7600,\\
Tour 24 - 2i\`{e}me \'et., 4 pl. Jussieu, F 75252 Paris Cedex 05, France}
\author{C.-L. Ho}
\email{ hcl@mail.tku.edu.tw}%
\affiliation{Department of Physics, Tamkang University, Tamsui
251, Taiwan, R.O.C.}
\author{G. H. E. Duchamp}
\email{ghed@lipn-univ.paris13.fr}
\affiliation{Universit\'e Paris 13, Sorbonne Paris Cit\'e,\\ 
Laboratoire d'Informatique de Paris-Nord (LIPN),\\ 
CNRS, UMR 7030,\\ 
F-93430, Villetaneuse, France.}

\date{Jan 18, 2013} 

\begin{abstract}

We propose a unified mathematical scheme,  based on a classical tensor isomorphism,
for characterizing entanglement that works for pure states of multipartite systems of
any number of particles.   The degree of entanglement is indicated by a set of
absolute values of the determinants for each subspace of the multipartite systems.
Our scheme provides a characterization of the degrees of entanglement when
the qubits are measured or lost successively, and leads naturally to
necessary and sufficient conditions for multipartite pure states to be separable.
For systems with a large number of particles, a  rougher indication of the degree of entanglement is
provided by the set of  mean values of
the determinantal values  for each subspace of the multipartite systems.

\end{abstract}





\maketitle

\section{Introduction}

As Schr\"odinger once said, ``...entanglement is {\em the} characteristic
trait of quantum mechanics, the one  that forces its entire departure from classical lines of thought. " \cite{schrod}.
Entanglement now lies at the heart of quantum
information.  As such, characterizing entanglement is currently one of the
main tasks in quantum information theory.

For bipartite states, it is convenient  that a single determinantal
condition is enough to discriminate between separability and
entanglement. This is the so-called concurrence \cite{wootters}.
Unfortunately, for multi-partite systems it seems impossible  to
classify the  degree of entanglement with a single quantity like
the concurrence.

In the literature, most people have been  interested in defining
measures of entanglement which are invariant under local unitary
transformations. This is motivated by the belief that one has the
freedom to measure the system in any direction, and that
entanglement should be invariant under such freedom of
measurement 
(for a general discussion of various entanglement measures, see e.g.: \cite{V}).

For tripartite states, a measure of entanglement invariant under
local unitary transformations was given in \cite{ckw}. This
measure, called the 3-tangle in \cite{ckw}, turns out to be just
the Cayley hyperdeterminant (Det) of the corresponding system.
Such tangle type formulation of entanglement measure has been generalised to $N$-qubit systems in \cite{OV}.

Unfortunately, the 3-tangle cannot be 
  a complete measure  
 of
entanglement for tripartite states.  One needs only mention the fact
that the Cayley hyperdeterminant for the well-known GHZ-state \cite{GHZ}
and
the W-state \cite{W} are one and zero, respectively. 
However, the W-state 
{\em is}  a genuinely entangled tripartite state;
 so ${\rm Det}=0$ does not provide a criterion
for separability as the simple concurrence does in the bipartite
case. Further, one knows that the W-state is in fact more robust
under measurement-collapse than the GHZ-state \cite{hso}. For
example, if Alice measures the (first) qubit of the GHZ-state to
be $0$, then this leaves  the separable state $\ket{00}$.  And
similarly for any measurement of any qubit in any of the three
subspaces for the tripartite GHZ-state. On the other hand, the
determination
({\em in the same basis})
 of the value ``0" of any qubit in any space for  the
W-state still leaves the state (maximally) entangled, and only if
the value ``1" is measured will the collapsed state be separable.
Again this difference is not reflected in the values of the Cayley
hyperdeterminant for these two states. So one needs additional
indicators to reflect this difference in entanglement properties.

Motivated by such observation, we propose to consider entanglement properties of  multipartite systems 
as each one of its  qubits is successively measured in the same basis, in order to supplement  other measures which 
concern mainly the invariant properties of entanglement under local unitary transformations without making any measurement, i.e., without losing any qubit.

One such scheme was presented in \cite{sh}, where six additional indicators (sub-determinants) were introduced to
supplement  the Cayley hyperdeterminant. 
Together these 7 numbers
distinguish the GHZ-state, the W-state and other tripartite
states, and - more significantly - provide a necessary and
sufficient criterion for the separability of a tripartite pure
state.
 Of these 7 numbers,  the Cayley hyperdeterminant is linked to the tripartite system,
and the six sub-determinants indicate degrees of entanglement of the 6 possible
bipartite systems when one of the three qubits is measured.

One would like to extend the scheme in \cite{sh} to multipartite
cases. Unfortunately,  it is not easy to generalize the Cayley
hyperdeterminant to higher dimensional systems (for a generalization to four qubits, see e.g., \cite{ost1}).  It is therefore
advantageous to find a classification scheme of
entanglement that applies to any number of particles.
There have been some attempts to characterize entanglement based on certain invariants under local unitary transformations (see e.g., \cite{ost2}).

In this note we propose a unified mathematical scheme that works
for multipartite systems of any number of particles. This scheme
is based on a classical tensor isomorphism, as
expounded - for example - by Bourbaki \cite{bourbaki}.
It provides an indication of the degrees of entanglement when
the qubits are measured or lost successively.

\section{Bipartite spaces, Concurrence}

 We first consider the property of the {\em separability} of
{\em bipartite   pure states}.
\begin{dex} [Separable state] 
{\bf 
(see e.g.: \cite{Wer})
}
An   element $v \in V_1 \otimes V_2$ is said to be {\em separable}
(equivalently, non-entangled) if $v$ can be written as a direct
product
$$v=v_1 \otimes v_2\; \; \; \; \; v_1 \in V_1, v_2 \in V_2.$$
\end{dex}
Note that this statement is basis-dependent.  For example, if
$V_1$ and $V_2$ are qubit spaces (that is, their elements are
complex 2-vectors of norm $1$), then by a suitable non-local unitary
transformation $U$ ($U \in U(4)$) given by 
\begin{eqnarray}
U=
 \left( \begin{array}{cccc}
  1 & 0 & 0 &0 \\
  0 & 1 & 0 & 0 \\
  0 & 0 & 0 & 1 \\
  0 & 0 & 1 & 0
  \end{array}\right),\label{CNOT}
\end{eqnarray}
the separable  state $(\alpha \ket {0}+ \beta \ket{1})\ket{0}$ is transformed to
\begin{equation}
   U(\alpha \ket {0}+ \beta \ket{1})\ket{0}= \alpha \ket{00} + \beta \ket{11},
   \end{equation}
which is non-separable insofar as $\alpha, \beta\neq 0$. 
The transformation $U$ in Eq.~(\ref{CNOT}), commonly called the CNOT-gate,  is non-local, and is used to generate entangled states in quantum computation.

 However, the property of being separable is
obviously conserved under a local unitary transformation
 (and even by local transformations)
defined as follows:

\begin{dex}[Local unitary transformation]
The transformation $U$ on $V_1 \otimes V_2$ is said to be a
local unitary transformation
 if $U=U_1 \otimes U_2$ where $U_i$ is a unitary operator acting on
$V_i$.
\end{dex}
Clearly the above  considerations hold for general direct product spaces.

For bipartite pure qubit states a single determinantal condition
is enough to discriminate between separability and entanglement.

Let $v \in V = V_1\otimes V_2$, where $V_1$ and $V_2$ are two-dimensional (qubit) vector
spaces with basis
\begin{equation}\label{2-basis}
   \{e_0 \equiv \ket{0} \equiv [1,0]^{T},e_1 \equiv \ket{1}\equiv [0,1]^{T}\}.
\end{equation}
In general we may write  $v \in V$ as
\begin{equation}\label{bip}
  v \in V_1\otimes V_2 =\sum_{i,j=0}^{1}c_{ij} \; \; e_i \otimes e_j\,.
\end{equation}

If $v$ is  separable, then
\begin{equation}
 v= (x_{0}e_{0}+x_{1}e_{1})\otimes(y_{0}e_{0}+y_{1}e_{1}),
\end{equation}
and so
\begin{equation}\label{bisep}
   c_{ij}=x_{i}y_{j} \;\;\;\; \{i,j=0,1\}
\end{equation}
from which we deduce that the matrix $c$ of coefficients $c_{ij}$
has determinant zero, $\det c=0$ or, equivalently, is of rank $1$.
This condition is clearly necessary and sufficient.

In fact, by suitably normalizing, we may use this determinant to
provide a {\em measure} of entanglement for pure states called the
{\em concurrence} ${\mathcal C}$, with
\begin{equation}\label{2-conc}
   {\mathcal C}=2|\det c|.
\end{equation}
This measure of entanglement varies between 0 (separable) and 1
(maximally entangled) and may be conveniently extended to mixed
states \cite{wootters}.  It may be shown by direct calculation
that this measure of entanglement is invariant under local unitary
transformations (see for example \cite{sh}).

For tripartite (and higher) spaces we shall show that a single
number is {\em not} sufficient to describe separability (or the
measure of entanglement).


\section{Finite Vector Spaces}

Since we shall be regarding  pure-state entanglement as
essentially a property of the elements of direct product vector
spaces, we briefly review some definitions.

\subsection{Direct Product Spaces}

We consider ${\mathcal V}$ a (finite) direct product of $m$
(finite) vector spaces:
\begin{equation}\label{dir}
{\mathcal V}=V_1 \otimes V_2 \otimes \ldots \otimes V_m
\end{equation}
where $V_r$ ($r=1,2,\ldots,m$) has basis
$$ \{e^{(r)}_0,e^{(r)}_1,\ldots,e^{(r)}_{n_r-1} \}$$
and the underlying field will be taken to be $\mathbb{C}$.
The  basis elements of the dual $V^{*(r)}$ are given by
$$ \{e^{*(r)}_0,e^{*(r)}_1,\ldots,e^{*(r)}_{n_r-1} \}$$
whose elements are defined by
$$\langle{e^{*(r)}_{i}},{e^{(r)}_{j}}\rangle = \delta_{ij},~~i,j=0,1,\ldots,n_r-1$$
using a standard notation for the action of a dual element on a vector.
The space $V$ has product basis
\begin{equation}\label{basis}
\{e^{(1)}_{s_1} \otimes e^{(2)}_{s_2} \otimes \ldots  \otimes
e^{(m)}_{s_m}| s_i=0 \ldots n_i-1 \}
\end{equation}
The dual space ${\mathcal V}^{*}$ also has dimension $\Pi_{r=1}^{m}{n_r}$, with the standard dual basis
\begin{equation}\label{dualbasis}
    \{e^{*(1)}_{s_1} \otimes e^{*(2)}_{s_2} \otimes \ldots  \otimes e^{*(m)}_{s_m}| s_i=0 \ldots n_i-1\}.
\end{equation}

\subsection{Rank of an element of a direct product space}

As a preliminary we consider the direct product of {\em two}
spaces, $E={\mathbb{C}}^2$, $F={\mathbb{C}}^2$.  For $u \in E
\otimes F$ we have $u=\Sigma_{i}\;{x_i \otimes y_i}\; \; \; (x_i
\in E, y_i \in F)$.  Following \cite{bourbaki}, we may define a linear map $u_1$ corresponding
to $u$ by
\begin{eqnarray}\label{u1}
   u_1: E^* &\rightarrow& F \\ \nonumber
        x^* &\mapsto& \Sigma_{i}{\langle x^{*},x_i \rangle y_i}.
\end{eqnarray}

\begin{dex}[Rank of an element of a product space] {\rm \cite{bourbaki}}
The Rank of $u \in E \otimes F$ is defined as the rank of $u_1$ (as a linear map).
\end{dex}

\begin{example}[Bipartite qubit state]
{\rm
 Let $u \in V \otimes V$ where $V$ is the qubit space
$\mathbb{C}^2$. Then $u= \Sigma_{i}x_i \otimes y_i$, and the
corresponding linear map $u_1:V^{*}\rightarrow V$ is given by
\begin{equation}
   u_{1}(v^{*}) = \Sigma_{i}{\langle v^{*}, x_i \rangle y_i} \; \; \; \; (x_i, y_i \in V)
\label{u1-v}
\end{equation}

If $u$ consists of a single product, then choosing a basis,
\begin{equation}
u=x \otimes y \; \; \; \; x=\left[\begin{array}{c} a \\ b
\end{array} \right]\; \; \; \;y=\left[\begin{array}{c} c\\ d
\end{array} \right].
\end{equation}
Now for any $v\in V$ given by
\[
v=\left[\begin{array}{c} v_1 \\ v_1
\end{array} \right],
\]
the action of the map $u_1$ in (\ref{u1-v}) is
\begin{eqnarray}
   u_{1}(v^*)
   & =&\left([v_1,v_2]\,\left[\begin{array}{c} a \\ b
\end{array} \right]\right)\,\left[\begin{array}{c} c \\ d
\end{array} \right]  \nonumber\\
    &=& \left[\begin{array}{cc} ac & bc\\ad & bd \end{array} \right]\left[\begin{array}{c} v_1 \\ v_1
\end{array} \right],\nonumber
\end{eqnarray}
Hence
\begin{equation}\label{u1-2-qubit}
    u_1=\left[\begin{array}{cc} ac & bc\\ad & bd \end{array} \right],
\end{equation}
which has Rank 1.
 Conversely, if $u=\Sigma _{i}(x_i \otimes y_i),
\; \; \; x_i=a_i e_0 + b_i e_1 , \;y_i=c_i e_0 + d_i e_1 $, then
\begin{equation}\label{uu-2-qubit}
    u=AC e_0 \otimes e_0 +AD e_0 \otimes e_1 +BC e_1 \otimes e_0 +BD e_1 \otimes e_1,
\end{equation}
with
$$
AC \equiv \Sigma_{i}{a_i c_i} \; \;AD \equiv \Sigma_{i}{a_i d_i}
\; \;BC \equiv \Sigma_{i}{b_i c_i} \; \;BD \equiv \Sigma_{i}{b_i
d_i}
$$
and
\begin{equation}\label{uu1-2-qubit}
    u_1=\left[\begin{array}{cc} AC & BC\\ AD & BD  \end{array} \right],
\end{equation}
where not all $AC, AD,BC, BD$ are zero.  Without loss of
generality we may choose $AC \neq 0$.\\
Thus $\det(u_1)=0
\Rightarrow u=(AC e_0+BC e_1)\otimes(e_0+\frac{AD}{AC}e_1)$ and
$u$ is separable. }
\end{example}

We therefore have the following theorem:
\begin{theorem}
A necessary and sufficient condition for the bipartite vector $u
\in V \otimes V$ to be separable is that the corresponding linear
transformation $u_1$ be of Rank 1.
\end{theorem}

\subsection{Rank of an element of a multi-direct product space}\label{rankmp}

We now extend the result of the previous section to a multi-direct
product space. We consider the multipartite vector space
${\mathcal V}$ of Eq.(\ref{dir}):
\begin{equation*}
{\mathcal V}=V_1 \otimes V_2 \otimes \ldots \otimes V_m
\end{equation*}
and remark that the space of homomorphisms from
$V_1 \otimes V_2 \otimes \ldots \otimes V_{m-1}$ to $V_m$ is isomorphic to\\
$(V_1 \otimes V_2 \otimes \ldots \otimes V_{m-1})^{*}\otimes V_m$; that is,
\begin{eqnarray}\label{hom1}
{\mathcal{H}}om(V_1 \otimes V_2 \otimes \ldots \otimes V_{m-1},V_{m})
 &\simeq&  (V_1 \otimes V_2 \otimes \ldots \otimes V_{m-1})^{*}\otimes V_m \nonumber\\
               &\simeq& V_1 \otimes V_2 \otimes \ldots \otimes V_{m-1} \otimes V_{m} \label{hom2}.
\end{eqnarray}
Corresponding to an element
$$u \in {\mathcal V},\; \; \;  u=\Sigma_i{w_i \otimes z_i},\;\;\; (w_i \in V_1 \otimes V_2 \otimes \ldots \otimes V_{m-1}, z_i \in V_m)$$
we define the linear map
\begin{eqnarray}\label{lin}
    u_1:V^*_1 \otimes V^*_2 \otimes \ldots \otimes V^*_{m-1}& \rightarrow & V_m\nonumber \\
    v^{*} & \mapsto & \Sigma_{i}{\langle v^{*}, w_i \rangle z_i}.
\end{eqnarray}

We may therefore define the {\em Rank} of the direct product $u$  by the rank of the linear map $u_1$ in Eq.(\ref{lin}).
\\ \\
For other than bipartite states there are of course varying degrees of separability.  We propose the following definition:
 \begin{dex}[Partial separability of an element of a product space]\label{partial}
 We say that the vector $u$
\begin{equation*}
u \in {\mathcal V}=V_1 \otimes V_2 \otimes \ldots \otimes V_m
\end{equation*}
is partially separable with respect to $V_s \; \; (1\leq s\leq m)$ if
 \begin{equation*}
u=v\otimes z \otimes w \;\;\;\;\;(v \in V_1 \otimes V_2 \otimes \ldots \otimes
V_{s-1},\; z \in V_s, w \in V_{s+1} \otimes V_{s+2} \otimes \ldots \otimes V_{m}).
\end{equation*}
 \end{dex}

 We may then also define:
  \begin{dex} [Complete separability of an element of a product space]\label{complete}
 The vector $u$
\begin{equation*}
u \in {\mathcal V}=V_1 \otimes V_2 \otimes \ldots \otimes V_m
\end{equation*}
is completely separable  if it is partially separable with respect to $V_s, s=1 \ldots m$.
\end{dex}
We shall often simply say in the foregoing case that $u$ is separable.

It therefore follows from the preceding discussion that a necessary and sufficient condition for $u$ to be partially separable
with respect to  $V_m$ is that $u_1$ be of rank one.
When this condition is fulfilled we may write $u=w\otimes z$ ($z$ is taken in the dimension 1 image of $u_1$ and the
decomposition is then unique up to scalars).

 For {\em complete separability} of $u$  as defined above we must demand that $w$  be completely separable, and so on recursively.
 This procedure then provides
  not only a necessary and sufficient condition for $u$ to be (completely) separable but gives an algorithm for its  factorization.

\section{Tripartite states}

We now show how the ideas discussed in Sect.~III provide indicators of degrees of entanglement in  tripartite systems.

\subsection{Separable cases}

As an important example, we consider a (completely) separable element  of a tripartite
qubit space.

We take a simple direct product tripartite state (separable
pure state)
\begin{equation}\label{sep1}
    u\in V_1 \otimes V_2 \otimes V_3 \; \; \; \; (V_i \equiv \mathbb{C}^2)
\end{equation}
with
\begin{equation}\label{sep2}
    u=w\otimes z, \; \; \; (w = x\otimes y \in V_1 \otimes V_2, \; \; \; z \in V_3).
\end{equation}

According to the foregoing discussion, to $u$ corresponds  the linear map
\begin{eqnarray}
 u_1:V^*_1 \otimes V^*_2 &\rightarrow& V_3 \nonumber\\
    v* &\mapsto& {\langle v^{*}, w \rangle z} \; \; \; \; .
    \end{eqnarray}
 Choosing a basis,
\begin{equation}
w=x \otimes y \; \; \; \; x=\left[\begin{array}{c} x_0 \\ x_1
\end{array} \right]\; \; \; \;y=\left[\begin{array}{c} y_0\\ y_1
\end{array} \right]\; \; \; \;z=\left[\begin{array}{c} z_0\\z_1
\end{array} \right]
\end{equation}
and so
\begin{equation}\label{u1-3-qubit}
u_1 =\left[\begin{array}{cccc}
x_0y_0z_0 & x_0y_1z_0 & x_1y_0z_0 & x_1y_1z_0\\
x_0y_0z_1 & x_0y_1z_1 & x_1y_0z_1 & x_1y_1z_1
\end{array} \right]
\end{equation}
which has Rank 1.
Eq.~(\ref{u1-3-qubit}) is a direct extension  of (\ref{u1-2-qubit}).
Note that if we replace the term $w \otimes z$ by $\Sigma_i{w_i}
\otimes z$ the Rank is still $1$.  This shows that Rank=$1$ is
necessary but not sufficient for (complete) separability. This preliminary condition on $u_1$ guarantees
partial separability with respect to $V_3$, as in Definition \ref{partial}. For complete
separability as defined in Definition \ref{complete} we require the  further condition that the Rank of
the element of $V_1 \otimes V_2$ also be $1$.

\subsection{General cases}

We now discuss the important example of a general tripartite qubit
state.  We follow the procedure above in \ref{rankmp} for the
general multipartite case, which it illustrates.

 Consider  a general element $u$ of a tripartite qubit space
\begin{equation}\label{gts}
    u\in V_1 \otimes V_2 \otimes V_3 \; \; \; \; (V_i \equiv \mathbb{C}^2).
\end{equation}
We  choose the standard basis as in Eq.(\ref{2-basis}), to write
\begin{eqnarray}\label{gentri}
  u&=& \sum_{i,j,k=0}^{1} {x_{ijk}\ket{ijk} }\nonumber\\
   &=&  \sum_{i,j,k=0}^{1} {x_{ijk} \; e_i \otimes e_j \otimes e_k} \\
   &=&\sum_{k=0,1}{w_k \otimes z_k }\; \; \; \; (w_k \in V_1 \otimes V_2 ,\; \; z_k \in V_3)\nonumber
   \end{eqnarray}
with $w_0 =\sum_{ij}{x_{ij0} \;e_i \otimes e_j},\; \; w_1 =\sum_{ij}{x_{ij1}\; e_i \otimes e_j},\; \; z_k=e_k$.
As in Eq.(\ref{lin}) above, we  define the linear map
\begin{eqnarray}\label{lin2}
    u_1:V^*_1 \otimes V^*_2  &\rightarrow&  V_3 \nonumber\\
    v^{*} & \mapsto & \sum_{k}{\langle v^{*}, w_k \rangle z_k}.
\end{eqnarray}
Writing in the standard basis
$$v^{*} = \sum_{m,n=0}^1 {v_{mn}\; e^{*}_m \otimes e^{*}_n }= \left( v_{00},v_{01},v_{10},v_{11}\right)$$
the relevant linear transformation in matrix form is
\begin{equation}\label{matrix}
    \left( v_{00},v_{01},v_{10},v_{11}\right) \mapsto \left[\begin{array}{cccc}x_{000}&x_{010}&x_{100}&x_{110}\\
    x_{001}&x_{011}&x_{101}&x_{111}\end{array} \right]\left( v_{00},v_{01},v_{10},v_{11}\right)^{T}.
\end{equation}
The condition for separability between the $V_1 \otimes V_2$ space
and $V_3$ is that the $2 \times 4$ matrix in Eq.(\ref{matrix})
should be of rank 1.
 This means, from Eq.~(\ref{matrix}), that the following six determinants
\begin{eqnarray}\label{dets}
\det(1)_3&=& x_{{000}}x_{{011}}-x_{{001}}x_{{010}},\nonumber \\
\det(2)_3&=& x_{{000}}x_{{101}}-x_{{001}}x_{{100}} ,\nonumber \\
\det(3)_3&=& x_{{000}}x_{{111}}-x_{{001}}x_{{110}} ,\nonumber \\
\det(4)_3&=& x_{{010}}x_{{101}}-x_{{011}}x_{{100}},  \\
\det(5)_3&=& x_{{010}}x_{{111}}-x_{{011}}x_{{110}}, \nonumber \\
{\rm  and} ~~~~\det(6)_3&=& x_{{100}}x_{{111}}-x_{{101}}x_{{110}}\nonumber
\end{eqnarray}
are identically zero.
Here the subscript ``3" indicates that these determinants are related to the tripartite system.
The vanishing of these 6 determinants essentially  means that the vectors $x_{ij0}$ and
$x_{ij1}$ are parallel, or either one of them is a null vector.
Complete separability is then attained by applying the Rank 1
condition to the 4-vector $[x_{000},x_{010},x_{100},x_{110}]$
(or to $[x_{001},x_{011},x_{101},x_{111}]$), namely that the
determinant (subscript ``2" indicates that the determinant is related to the bipartite system)
\begin{equation}\label{det2}
   \det(1)_2=\left|\begin{array}{cc}x_{000}&x_{100}\\x_{010}&x_{110}\end{array}\right| ~~ {\rm~ or}~~
   \det(2)_2=\left|\begin{array}{cc}x_{001}&x_{101}\\x_{011}&x_{111}\end{array}\right|
\end{equation}
vanishes.

In the case where the 6 determinants in Eq.~(\ref{dets}) are {\em not} identically zero, the third qubit is entangled with the other two.
Then the two determinants in Eq.~(\ref{det2}) provide an indicator as to whether the remaining two qubits are entangled when the third qubit is lost:
$\det(1)_2=0$ ($\det(2)_2=0$) means the two qubits are separable if the third qubit is measured to be ``$0$" (``$1$").

The above discussion then suggests a classification scheme of entanglement by a list of the 8 determinants
given by
$$\left[|\det(1)_3|,|\det(2)_3|, |\det(3)_3|, |\det(4)_3|, |\det(5)_3|, |\det(6)_3|; |\det(1)_2|,|\det(2)_2|\right] .$$
One may define a set of  coarse-grained indicators of the tripartite  entanglement  by
\begin{eqnarray}
\left[\mathcal{C}_3,\mathcal{C}_2 \right], ~~~
\mathcal{C}_m=\frac{1}{l_m}\sum_{k=1}^{l_m}\,|\det(k)_m|,
\label{coarse}
\end{eqnarray}
where $l_3=6, l_2=2$ are the numbers of determinants for the 3- and 2-partite spaces.

In Table \ref{tristates} we present this  classification scheme for
some tripartite states based on the methods of this note.
As in \cite{sh}, the determinants are all normalized to
$1$ by applying a
normalization factor $1/|{\rm det} A|$ for all non-vanishing determinants.

\begin{table}[h]
\centering \caption {\label{tristates}  Classification of some
tripartite qubit states.}
\begin{tabular}{@{}l*{15}{c}*{15}{c}*{15}{c}*{15}{c}}
\hline
\hline
& State &  $[|\det(k)_3|,~k=1,2,\ldots,6]$ & $[|\det(k)_2|,~k=1,2]$ & $[\mathcal{C}_3$; ${\cal C}_2]$\\
\hline
General Separable State & $\Sigma a_ie_i\otimes\Sigma b_je_j\otimes\Sigma c_ke_k$ & [0,0,0,0,0,0] & [0,0] &[0;0]\\
GHZ-state & $(1/\sqrt{2})(\ket{000}+\ket{111})$ & [0,0,1,0,0,0] & [0,0] & [$\frac{1}{6}; 0$]\\
W-state &$(1/\sqrt{3})(\ket{001}+\ket{010}+\ket{100})$ & [1,1,0,0,0,0] & [1,0] & [$\frac{1}{3};\frac12$]\\
Cluster state &
$(1/\sqrt{8})(\ket{000}+\ket{001}+\ket{010}-\ket{011}$ &
[1,0,1,1,0,1] & [1,1] & [$\frac{2}{3}; 1$]\\
& $+\ket{100}+\ket{101}-\ket{110}+\ket{111})$ & &\\
$\psi$-state  \cite{sh}& $(1/2)(\ket{001}+\ket{010}+\ket{100}+\ket{111})$ & [1,1,0,0,1,1] & [1,1] & [$\frac{2}{3}; 1$]\\\
$\phi$-State \cite{sh} &$(1/2)(\ket{000}+\ket{011}+\ket{101}+\ket{110}$ & [1,1,0,0,1,1] & [1,1] & [$\frac{2}{3}; 1$]\ \\
\hline \hline
\end{tabular} 
\end{table}

From the table it is seen that $\det(k)_3$ are not identically zero for the GHZ, W, $\psi$, $\phi$ and the cluster state,
meaning that the three qubits are entangled for these states. Yet from the values of $\det(k)_2$, it is clear that the GHZ state
is completely separable when it loses its third qubits, while the W-state is only separable if the third qubit is measured to be ``1".
The $\psi$ and $\phi$ states are obtained from the GHZ state by local unitary transformations \cite{sh}, yet unlike the GHZ state, 
they remain entangled whatever the measured value 
(in the $\{|0\rangle,|1\rangle\}$ basis)
is for the third qubit.  Note that these two
states are in fact identical, related simply by a redefinition of $|1\rangle$ and $|0 \rangle$,which the classification scheme  exposes.

These eight  determinants thus give us a complete picture of the degrees of entanglement of the tripartite systems, while
the two coarse-grained indicators $\mathcal{C}_3$ and $\mathcal{C}_2$ provide a rough idea of the degrees of entanglement of the tripartite
systems and the reduced bipartite systems when the 3rd qubit is lost.

\subsection{Cayley Hyperdeterminant}

In  \cite{sh} it was also shown that 7 parameters
were necessary in order to determine the separability of a
tripartite qubit state. The analysis  involved showing that 6
submatrices had Rank 1, as well as the vanishing of  the Cayley
hyperdeterminant.  This was shown to give a necessary andy
sufficient condition for separability.  For the state in
Eq.(\ref{gentri}), the six submatrices of \cite{sh} were given by
\begin{eqnarray}\label{submat}
A_{x_0}&=& \left(x_{0ij}\right),~~
A_{x_1} =  \left(x_{1ij}\right)~~ \\
A_{y_0}&=& \left(x_{i0j}\right),~~
A_{y_1} =  \left(x_{i1j}\right)~~\nonumber \\
A_{z_0}&=& \left(x_{ij0}\right),~~
A_{z_1} =  \left(x_{ij1}\right)\nonumber
\end{eqnarray}
and the Cayley Hyperdeterminant ${\rm Det} A$ is given
by\footnote{We take this opportunity to correct a typographic error
in \cite{sh}, where the third term in Eq.~(\ref{cayley}) was
erroneously left out.}
\begin{eqnarray}\label{cayley}
{\rm Det}\, A &=& x_{000}^2 x_{111}^2 + x_{001}^2 x_{110}^2 +
x_{010}^2 x_{101}^2 + x_{100}^2 x_{011}^2\nonumber\\
 &-& 2\left[x_{000} x_{001} x_{110} x_{111} + x_{000} x_{010} x_{101} x_{111}
 + x_{000} x_{011} x_{100} x_{111}\right.\nonumber\\
 &+&\left.x_{001} x_{010} x_{101} x_{110}
 + x_{001} x_{011} x_{101} x_{100} +  x_{010} x_{011} x_{101} x_{100}
 \right]\label{Det-x}\\
 &+& 4\left[x_{000} x_{011} x_{101} x_{110} +  x_{001} x_{010} x_{100}
 x_{111}\right].\nonumber
\end{eqnarray}

Of these 7 numbers,  the Cayley hyperdeterminant is linked to the tripartite system,
and the six sub-determinants indicate degrees of entanglement of the 6 possible
bipartite systems when one of the three qubits is measured.

The submatrices Eq.(\ref{submat})  are {\em not} the same as the
submatrices considered in this paper, namely, Eq.~(\ref{dets}).
Specifically,  the Cayley hyperdeterminant is in fact a function
of the determinants Eq.(\ref{dets}) considered here, namely,
\begin{equation}
{\rm Det} A
=\det(3)^{2}+\det(4)^{2}-2\,\det(2)\det(5)-2\,\det(1)\det(6).
\end{equation}
One advantage of the previous approach in \cite{sh} is that the
determinants of the submatrices (\ref{submat}) have a physical
significance, being the subconcurrences, as has the Cayley
Hyperdeterminant which in \cite{ckw} was considered as the
3-tangle.  However, it is not easy to see how the Cayley
Hyperdeterminant may be generalized to higher multipartite
systems, and the current approach used here appears more direct.

\section{The $N$-qubit case}

It is now clear how the recipe elucidated in the previous section can be directly extended to $N$-qubit systems.

To determine the separability of the  $N$-qubit state
\begin{equation}\label{nqs}
    u\in V_1 \otimes V_2 \otimes \ldots \otimes V_N \; \; \; \; (V_i \equiv \mathbb{C}^2).
\end{equation}
the outlined procedure involves determining first the rank of $u_1$
\begin{equation}\label{linn}
    u_1:V^*_1 \otimes V^*_2 \otimes \ldots V^*_{N-1} \rightarrow  V_N.
\end{equation}
In the standard basis, $u_1$ is represented by a $2 \times 2^{N-1}$ matrix.

Partial separability with respect to
$V_N$ is guaranteed by the rank of $u_1$ being $1$; i.e. the $\binom{2^{N-1}}{2}$ $2 \times 2$ submatrices of
$u_1$ must have determinant zero,
$\det(k)_N=0,~~k=1,2,\ldots,
\binom{2^{N-1}}{2}$.

Proceeding recursively,   one sees that the degrees of entanglement of a $N$-qubit system are indicated
by the list of determinants
\begin{eqnarray}
\left[\{\det(k)_m, ~k=1,2,\ldots,l_m\};~m=N, N-1, \ldots,2\right].
\end{eqnarray}
Here the number $l_m$ of determinants for $\{\det(k)_m\}$ is
\begin{eqnarray}
 l_m=
  2^{N-m}\times \binom{2^{m-1}}{2}, & m=N,N-1,\ldots,2.
  \label{lm}
\end{eqnarray}
Clearly,
we have to
examine
\begin{eqnarray}
  \sum_{m=2}^N \frac{l_m}{2^{N-m}}=\; \binom{2^{N-1}}{2} +  \binom{2^{N-2}}{2}  + \cdots +\binom{2}{2}
 \end{eqnarray}
$2 \times 2$ submatrices to determine complete separability.
The factor $1/2^{N-m}$ for $l_m$ comes from the fact that when the $(m+1)$-th qubit is factorizable  from the rest of the $m$ qubits, then
the  remaining $m$-qubit is only of dimension $l_m/2$, because  either its coefficients associated with the $|0\rangle_m$ and  $|1\rangle_m$ are proportional , or
one set of them is identically zero.
This set of numbers gives
rise to the combinatorial sequence $1, 7, 35, 155, 651, 2667,
10795, 43435, 174251, 698027 \ldots$ which is, {\it inter alia},
the Gaussian Binomial Coefficient $[N,2]_{q=2}$ \cite{sloane} but in any case diverges exponentially.

Just as for the tripartite cases, one may define a set of  coarse-grained indicators for multipartite entanglement  by
\begin{eqnarray}
\left[\{\mathcal{C}_m\};~m=N, N-1, \ldots,2\right],
\end{eqnarray}
where $\mathcal{C}_m$ are as defined by Eq.~(\ref{coarse}).
We emphasize that $\mathcal{C}_m=0$ implies that the multipartite system, after losing $(N-m)$ of its qubits,  is separable in one of its remaining $m$ bits.
Hence $\mathcal{C}_m=0$ for all $m=N,N-1,\ldots,2$ means the multipartite system is separable.

We present in Table \ref{4-states} this classification scheme for some representative 4-qubit systems \cite{4qubit}, 
which are the generalization of the corresponding tripartite states in Table I.
From (\ref{lm}) one sees that there are  28, 12 and 4 (normalized) determinants to compute at the 4-, 3- and 2-qubit level, respectively. 
For simplicity of presentation we shall use the notation 
$[0_k,1_{28-k}]$ to indicate the values of the 28  determinants at the 4-qubit level, i.e.: $k$ ``0" and (28-k) ``1".
We note that the coarse-grained
indicators $[\mathcal{C}_4;\mathcal{C}_3; \mathcal{C}_2]$
for the generalized GHZ-state
 $(\ket{0000}+\ket{1111})/\sqrt{2}$ and  the W-state $(\ket{0001}+\ket{0010}+\ket{0100}+\ket{1000})/\sqrt{4}$ are, respectively,
 $[1/28;0;0]$ and $[3/28;1/6;1/4]$.  This means the W-state is more robust than the GHZ-state, and the fact that $\mathcal{C}_3=\mathcal{C}_2=0$
 for the GHZ-state indicates that the GHZ-state is completely separable when it loses one of its qubits.  Again, as in the tripartite cases, 
 the generalized $\psi$ and $\phi$ states, though related to the GHZ state by local unitary transformations,  remain entangled
 whenever one qubit is measured
(in the $\{|0\rangle,|1\rangle\}$ basis).

\begin{table}[h]
\centering \caption {\label{4-states}  Classification of some
4-qubit states.}
\begin{tabular}{@{}l*{15}{c}*{15}{c}*{15}{c}*{15}{c}*{15}{c}}
\hline
\hline
& State & $[|\det(k)_4|]$ &  $[|\det(k)_3|]$ & $[|\det(k)_2|]$ & $[\mathcal{C}_4$; $\mathcal{C}_3$; $\mathcal{C}_2]$\\
\hline
Separable & $\Sigma a_i e_i\otimes\Sigma b_j e_j\otimes\Sigma c_k e_k \otimes\Sigma d_l e_l$ 
& [$0_{28}$]  & [0,0,0,0,0,0, 0,0,0,0,0,0] & [0,0, 0,0] & [0;0;0]\\
GHZ & $(1/\sqrt{2})(\ket{0000}+\ket{1111})$ 
& [$0_{27},1$] & [0,0,0,0,0,0,0,0,0,0,0,0] & [0,0,0,0] & [$\frac{1}{28}; 0; 0$]\\
W &$(1/\sqrt{4})(\ket{0001}+\ket{0010}+\ket{0100}+\ket{1000})$ 
& [$0_{25},1_3$] & [1,1,0,0,0,0,0,0,0,0,0,0] & [1,0,0,0] & [$\frac{3}{28};\frac{1}{6};\frac14$]\\
Cluster \cite{cluster} &
$(1/\sqrt{4})(\ket{0000}+\ket{0011}+\ket{1100}-\ket{1111}$ 
& [$0_{24}, 1_4$] & [0,0,0,0,0,0,0,0,0,0,0,0] & [1,0,0,1] & [$\frac{1}{12}; 0;1$]\\
$\psi$ & $(1/\sqrt{8})(\ket{0001}+\ket{0010}+\ket{0100}+\ket{0111})$
& [$0_{12}, 1_{16}$]  & [1,1,0,0,1,1,1,1,0,0,1,1] & [1,1,1,1] & [$\frac{4}{7}; \frac{2}{3}; 1$]\ \\
& $+\ket{1000}+\ket{1011}-\ket{1101}+\ket{1110})$ & & &\\%
$\phi$ & $(1/\sqrt{8})(\ket{0000}+\ket{0011}+\ket{0101}+\ket{0110})$
& [$0_{12}, 1_{16}$]  & [1,1,0,0,1,1,1,1,0,0,1,1] & [1,1,1,1] & [$\frac{4}{7}; \frac{2}{3}; 1$]\ \\
& $+\ket{1001}+\ket{1010}-\ket{1100}+\ket{1111})$ & & &
\\ \hline \hline
\end{tabular}
 \end{table}
 

\section{Generalization to higher spin cases}

Our result in Sect. III.C  may be generalized to multipartite pure states of arbitrary dimension, i.e., to multi-qudit case with higher spins.
One simply replaces the qubit space $V_k$ 
in ${\mathcal V}=V_1 \otimes V_2 \otimes \ldots \otimes V_m$ by the corresponding qudit space.

Suppose we denote by $M$ the dimension of the qudit space with $M$ levels, $M\times N$ the bipartite space where the first and the second particle
have spin $M$ and $N$ respectively, and so on. Hence $M=2$ for a qubit, and $2\times 2$  for the state space of a two-partite qubit states, etc. 

Consider, for instance, a bipartite qutrit states (i.e., $3\times 3$ case)
\begin{equation}
u=\sum_{i,j=0}^{2} {x_{ij}\ket{ij} }.
\end{equation}
The necessary and sufficient condition for  its separability is that the matrix
\begin{eqnarray}
\left[\begin{array}{ccc}
x_{00}&x_{10}&x_{20}\\
x_{01}&x_{11}&x_{21}\\
x_{02}&x_{12}&x_{22}
 \end{array} \right]
 \label{qutrit}
\end{eqnarray}
has rank one.
Similarly, for a tripartite qutrit ($3\times 3\times 3$) states, the condition for partial separability of the third qutrit with the first two  qutrits is given by the rank=1 condition of the matrix
\begin{eqnarray}
 \left[\begin{array}{cccccc}
 x_{000} & x_{010} & \cdots & \cdots & x_{210} & x_{220}\\
  x_{001} & x_{011} & \cdots & \cdots & x_{211}  & x_{221}\\
 x_{002} & x_{012} & \cdots & \cdots & x_{212}& x_{222}
  \end{array} \right].
  \label{qutrit2}
\end{eqnarray}
When the third qutrit collapses to one of its three states after a measurement is made on it, the separability of the remaining bipartite qutrits is then determined as in bipartite case, with the matrix (\ref{qutrit}) formed from the first, the second, or the third row of Eq. ~(\ref{qutrit2}), according to wether the measured value of the third qutrit is 0, 1, or 2, respectively. 

As another example, consider the case with $M\times N\times L$ space. 
The  condition for partial separability of the third qudit with the first two  qudits is given by the rank=1 condition of the matrix
\begin{eqnarray}
 \left[\begin{array}{cccc}
 x_{000}  & \cdots & \cdots & x_{M-1,N-1,0} \\
  x_{001}  & \cdots & \cdots & x_{M-1,N-1,1} \\
\vdots & \vdots & \vdots & \vdots \\
 x_{0,0,L-1}  & \cdots & \cdots & x_{M-1,N-1,L-1} 
  \end{array} \right].
  \end{eqnarray}
And the separability of the remaining qudits is determined recursively as discussed before.

The procedure can be easily extended to all multi-qudit states.  Certainly, the number of sub-determinants to be computed increases as the number of particles and the dimension increase. This is unavoidable in any scheme of measure of entanglement. Our scheme has the advantage that the same prescription applies to the determination of separability of pure states with  any number of particles and  spin .

\section{discussion}

In this note we have proposed a unified mathematical scheme for
characterizing entanglement that works for pure states (vectors) of multipartite systems of
any number of particles. This scheme is based on Bourbaki's
approach of defining the rank of a vector  in terms of the associated linear  mapping.

Our scheme provides an indication of the degrees of entanglement when
the qubits are measured or lost successively.
A rougher characterization of the degree of entanglement is provided by a set of coarse-grained indicators defined by the mean values of
the absolute values of the determinants for each subspace of the multipartite systems.

For an $N$-qubit system, the
 number of parameters required to distinguish separability of the
state is equal to the Gaussian Binomial Coefficient.  This number
is, unfortunately, exponentially large for large $N$. Thus the
corresponding set of indicators cannot be taken as a practical
measure of multipartite entanglement.  However, it should be noted
that as yet there is no simple scheme to quantify multipartite
entanglement, and we believe that the present scheme is the most
systematic and mathematically  direct one.

In the tripartite case, the present scheme
requires 7 $2 \times 2$ determinants
to determine complete separability. These 7 numbers are
different from those given in \cite{sh}. We have shown that the
Cayley hyperdeterminant is expressible in terms of 6 of the 7
determinants presented here.

It is well known that the GHZ, $\psi$ and $\phi$ states are related by local unitary transformations,
and hence it is often said that they are {\em equivalent} because of this fact.
Yet, there is no denying that they behave differently when one qubit is lost, as we pointed out in the Introduction.
When, say, the first qubit is measured in the z-basis, GHZ becomes separable, while the other two states remain entangled.
We believe this has indeed been  noted by many people, though not always mentioned in the literature.
Some people then argue that, when the first qubit is measured in the x-basis, say, the remaining two qubits of the GHZ are still entangled,
since when the first qubit is expressed in terms of the eigenstates $|0\rangle_x$ and $|1\rangle_x$ of the spin operator $S_x$, the GHZ state is expressed as
\begin{equation}
|GHZ\rangle  =\frac{1}{\sqrt{2}} \left[|0\rangle_x \left(|00\rangle_z+|11\rangle_z\right)
+[|1\rangle_x \left(|00\rangle_z-|11\rangle_z\right)\right].
\end{equation}
But then it is also easily verified that, under the same measurement in the x-basis, the remaining two qubits of the $\psi$ and $\phi$ states are separable.
For example, the $\phi$ state is given by
\begin{eqnarray}
\nonumber\\
|\phi\rangle &=&\frac{1}{2\sqrt{2}}\left[|0\rangle_x \left(|0\rangle_z+|1\rangle_z\right)\left(|0\rangle_z+|1\rangle_z\right)
+|1\rangle_x \left(|0\rangle_z-|1\rangle_z\right)\left(|0\rangle_z-|1\rangle_z\right)\right].\nonumber
\end{eqnarray}

All in all,  the $\psi$ and $\phi$ states behave differently from the GHZ state when a qubit is measured in the {\it same} basis.
Actually, the very fact that the GHZ state behaves differently when a qubit is measured in different bases is already an indication
that certain properties of entanglement cannot be invariant under local transformations.
We believe that one must also look at such non-invariant properties in order to have a better understanding of entanglement.

We emphasize here that our scheme differs from other measures of entanglement \cite{V} in that we consider entanglement properties of multipartite systems 
as each one of its  qubits is successively measured in the same basis, while other measures concern mainly the invariant properties of entanglement under local unitary transformations without making any measurement, i.e., without losing any qubit.

We are still far from having a definitive way to quantify
multipartite entanglement. In this regard, we believe that an
interesting direction is that of relating entanglement to the link
structures of knot theory \cite{sh1,kauff}.

\begin{acknowledgments}

This work is supported in part by the
National Science Council (NSC) of the Republic of China under
Grant NSC-99-2112-M-032-002-MY3 and  the National Center for Theoretical Sciences (North branch) of NSC (CLH),  and the Agence Nationale de
la Recherche (Paris, France) under Program No. ANR-08-BLAN-0243-2 and program PHC
POLONIUM ``QuantComb'' (GHED).

\end{acknowledgments}

\newpage


\end{document}